\documentclass[11pt,titlepage]{article}
\pdfoutput=1
\usepackage{setspace}
\usepackage{epsfig}
\usepackage{ae}
\usepackage{aecompl}
\usepackage{amsmath}
\usepackage{graphicx}
\usepackage[round,numbers,sort&compress]{natbib}
\usepackage{color}
\usepackage{soul}
\usepackage[normalem]{ulem}
\makeatletter
\def\bgcolour{\@ifnextchar[\bgcolour@i{\bgcolour@ii{}}}
\def\bgcolour@i[#1]{\bgcolour@ii{[#1]}}
\def\bgcolour@ii#1#2{%
  \bgroup%
  \markoverwith{%
    \hbox{%
      \begingroup%
      \kern+.1em%
      \color#1{#2}%
      \strut%
      \vrule\@width.3em%
      \kern+.0em%
      \endgroup%
    }%
  } \ULon%
}
\makeatother

\begin{document}

\title{Phenomenological Model and Phase Behavior of Saturated and
Unsaturated Lipids and Cholesterol}

\author{
G. Garb{\`e}s Putzel and M. Schick\\
Department of Physics\\ University of Washington,  Box
  351560, Seattle, WA 98195-1560}

\date{\today}
\maketitle
\begin{abstract} 
We present a phenomenological theory for the phase behavior of ternary
mixtures of cholesterol and saturated and unsaturated lipids, one which
describes both liquid and gel phases, and illuminates the mechanism of the
behavior. In a binary system of the lipids, the two phase separate when
the saturated chains are well ordered, as in the gel phase, simply due to
packing effects. In the liquid phase the saturated ones are not
sufficiently well ordered for separation to occur. The addition of
cholesterol, however, increases the saturated lipid order to the point
that phase separation is once again favorable. For the system above the
main chain transition of the saturated lipid, we can obtain phase diagrams in
which there is liquid-liquid phase separation in the ternary system but
not in any of the binary ones, while below that temperature we obtain the
more common phase diagram in which a gel phase, rich in saturated lipid,
appears in addition to the two liquid phases.
\end{abstract} 

\section{Introduction}
The hypothesis that the plasma membrane is not uniform but consists of
``rafts'' rich in saturated lipids and cholesterol which float in a
sea of unsaturated lipids remains an extremely exciting and
controversial one \cite{simons97,edidin03,munro03}. 
The extent to which one can define such entities in a
biological membrane  and the driving forces tending towards their
formation is unclear. Much more certain are the results of studies on
non-biological bilayer membranes with compositions designed to mimic
those of biological ones.
They consist of a high-melting point lipid, usually saturated, a
low-melting point lipid, usually unsaturated, and cholesterol 
 \cite{veatch05}. These
systems readily display coexisting liquid phases. One, the liquid
ordered (lo) phase \cite{ipsen87}, is  rich in
cholesterol and saturated lipids whose chains are relatively
well-ordered; the other, the liquid disordered (ld) phase, is
rich in unsaturated lipids whose chains are not so well-ordered 
\cite{dietrich01,veatch02,veatch05}. In
addition, at the temperatures of interest, 
there is usually a gel phase rich in saturated lipids with
chains which are very well ordered. This phase behavior leads
to the possibility that rafts in biological membranes 
are simply regions of one liquid phase
in coexistence with, and surrounded by, the other liquid. 
A second scenario, engendered 
by the occurrence of a line of
critical points in the system at which the two liquid phases become one, 
is that rafts are transient aggregates, manifestations of fluctuations 
which are particularly large near the critical line of a system which is 
effectively two-dimensional \cite{honerkamp-smith08}. A third 
interpretation is that such aggregates are the results of a line-active 
agent, perhaps cholesterol, which lowers the free energy of such 
aggregates and results in a liquid with much structure, as in a 
microemulsion of oil and water which is brought together by a 
surface-active agent.

It should be noted that not all ternary systems of lipids and cholesterol 
exhibit a gel and two liquid phases. Of particular interest is the system 
of diphytanoyl\-phosphatidyl\-choline (diPhyPC), dipalmitoyl\-phosphatidyl\-choline
(DPPC), 
and cholesterol \cite{veatch06} which, at a temperature above the 
main-chain transition of the DPPC, exhibits a region of liquid-liquid 
coexistence in the middle of the Gibbs triangle; that is, at these
temperatures none of the 
three binary systems undergo phase separation, whereas the ternary system 
does. This is unusual behavior in a liquid system, 
although other examples are cited in
ref. \cite{francis63}. It is to be contrasted
with that of most ternary liquid systems in which the existence of 
any phase separation is directly tied to the existence of such
separation in one of the binary systems. As a consequence only two of
the components are actually necessary to bring about the separation. A simple
example is the ternary system of oil, water, and surfactant which, for
modest concentrations of surfactant, displays two liquid phases, one
oil-rich, the other water-rich with the surfactant partitioning between
the two. Clearly these phases evolve from the ones resulting from 
separation of oil and water in the binary system. It is conceivable
that the existence of two liquid
phases in the ternary cholesterol,
saturated and unsaturated lipid system could be similarly directly tied to the
existence of liquid-liquid phase separation in the cholesterol, saturated
lipid system, a separation which had been reported
earlier \cite{vist90}. However there is now abundant experimental
evidence \cite{krivanek08} that such separation does not occur in this binary
system. The significance of the phase diagram of the 
diPhyPC, DPPC, cholesterol system is that the 
existence of liquid-liquid phase separation in such ternary systems
cannot be tied to phase separation in {\em any} of the binary systems,
and must therefore be a consequence of the presence of {\em all} 
of the three components. This conclusion begs for theoretical understanding.

The simplest approach would be to employ regular solution theory which
can, in fact, give rise to the kind of phase diagram described
above \cite{prigogine54,francis63,novak87}, but to do so requires an attractive interaction between
at least two of the components which is extremely large, and for the system of
interest, inexplicably so. Further, regular solution theory considers
only the concentration of each component. To employ it, therefore, would be
to ignore the many degrees of freedom of the lipid tails which are
certainly involved in the transition from the liquid to the gel phase,
and might be involved in the liquid-liquid transition. Thus one must
consider explicitly that some of the components are lipids.
 
There is surprisingly little theoretical work on the phase 
behavior of cholesterol and lipids, however, 
and most of that concerns binary systems 
\cite{ipsen87,komura04,radhakrishnan99,khelashvili05} which, as emphasized
above, cannot explain the origin of liquid-liquid separation in the
ternary systems of interest. 
Of the few papers on the phase behavior of 
ternary systems of saturated and unsaturated lipids and cholesterol 
\cite{radhakrishnan99,radhakrishnan05,elliott06}, only that of
Radhakrishnan and McConnell 
\cite{radhakrishnan05} has produced a phase diagram
exhibiting liquid-liquid coexistence in the ternary system 
without  phase separation in any of the three binary ones. 
Their theory posits that the 
cholesterol and saturated lipids form 
complexes which then interact repulsively 
with the unsaturated lipids. The origin of this repulsion is not
specified. In the explicit realization of this idea 
\cite{radhakrishnan05}, 
the complexes are treated as a distinct, fourth 
component in the system, and the free energy is assumed to be well
approximated by
regular solution theory.  The authors note that such complexes have
never been isolated in bilayers, but stress that they should be
considered as rapidly fluctuating entities. The evidence from simulation
for such transitory
complexation with a specific stoichiometry is mixed 
\cite{pandit04,pandit07,zhang07}.
Nevertheless, they emphasize the utility of the
concept, as clearly demonstrated by its ability to reproduce the unusual phase
diagram of the diPhyPC, DPPC, cholesterol
system. The theory does not describe the more usual ternary mixtures of
lipids and 
cholesterol which, in addition to two liquid phases, also display a gel phase. This
is probably due to the choice of order parameter of the saturated lipid, one which
they define as the fraction of saturated lipid found in complexes. In the
pure saturated-lipid system there can be no complexes, so this order
parameter always vanishes and cannot distinguish between the system's gel and liquid
phases.

Given the transitory nature of the posited complexes, it 
would be advantageous to formulate a theory which describes the system
only in terms of the original three components, and which would provide
further insight into the physics governing the system. In addition, one
would also like it to describe systems exhibiting a gel phase.  Lastly, it should
be able to describe the evolution of the normal phase behavior
into the more unusual one displayed by the diPhyPC, DPPC, cholesterol system.
   
It is the purpose of this paper to present such a theory, one which 
highlights what we believe to be the basic 
physics of these ternary systems. This is easily stated. Saturated and 
unsaturated lipids undergo phase separation when the former are 
sufficiently ordered, as they are in the gel phase. The reason for this is 
a simple packing effect: the presence of one or more kinks in an 
unsaturated chain prevents the efficient packing, and free energy 
reduction, of the ordered, saturated, chains. Hence the system lowers its 
free energy by separating into a gel phase rich in saturated lipids, and a 
liquid phase rich in unsaturated lipids. When the saturated and 
unsaturated lipids are in a liquid phase, the former are not sufficiently 
ordered to bring about a phase separation. The addition of cholesterol
to the liquid phase, 
however, tends to increase the order of the saturated lipids, as is well 
known \cite{leathes25,levine71,mcintosh78,hung07}. With the addition of
enough cholesterol, the saturated 
chains become sufficiently ordered that phase separation from the 
unsaturated chains becomes free energetically favorable once again. The 
theory produces phase diagrams which display two liquid phases and 
a gel, as well as those like that of the diPhyPC system in which there is only 
liquid-liquid coexistence, and that in the middle of the Gibbs triangle.
\section{The Model}
We consider a system of cholesterol, whose concentration is denoted $c$,
a high melting-temperature
lipid, henceforth referred to as ``saturated'', whose concentration is $s$, and a low
melting-temperature lipid, henceforth referred to as ``unsaturated'', 
with concentration $u=1-s-c.$
By describing the system in
terms of two independent concentrations, rather than three independent
areal densities, we are making the implicit simplifying 
assumption that the actual areal density of the system is not a crucial
parameter in the description of its phase behavior. We define an order
parameter, $\delta$, of the saturated lipid which is related to the
order of its chains \cite{komura04}. It will be largest in the gel
phase, and is non-zero in the liquid phases. 

We first assume the  system
to be at a temperature, $T$, sufficiently high that there are only liquid phases.  
We write the free energy per particle of the system, in units of $kT$,
in the form
\begin{eqnarray}
\label{fliq1}
{\tilde f}_{liq}(T,c,s,\delta)& =&{\tilde f}_{mix}+{\tilde
f}_{chain,liq}+{\tilde f}_{int,liq},\qquad {\rm where}\\
{\tilde f}_{mix}&=& c\ln c+s\ln s+u\ln u,               
\end{eqnarray}  
is the usual entropy of mixing. The second term describes the
interactions between the saturated lipids, and is therefore proportional
to $s^2$,
\begin{equation}
{\tilde f}_{chain,liq}=J_{ss}s^2[k_1(\delta-1)^2+(\delta-1)^4].
\end{equation}
Note that the strength of the interaction between saturated lipids
depends upon their configuration, as is to be expected. The information
about the configurations is encapsulated in the order parameter.
Because of this interaction between saturated lipids, the free energy in
the liquid phase has
a single minimum as a function of $\delta$. Taking advantage of the freedom to set the scale of the temperature and of the order parameter, we have used one of these degrees of freedom to assign $\delta$  the value unity in 
the pure saturated-lipid system. The constants $J_{ss}$ and $k_1$ are
positive.
The last term in eq. (\ref{fliq1}) is that due to the interaction between
the saturated lipid and the other two components,
\begin{equation}
{\tilde f}_{int,liq}=J_{us}us\delta-J_{cs}cs(\delta-k_2\delta^2),
\end{equation}
with $J_{us},\ J_{cs}$ and $k_2$ all positive.
The first term represents the repulsive interaction between saturated and
unsaturated lipids due to the inability of the latter to pack well with
the former. The strength of this repulsion depends, again, on the degree
of order of the chains of the saturated lipids, a concept expressed in
the earliest modeling of these systems \cite{ipsen87}. It is this term
which will drive the separation of saturated and unsaturated chains if
its strength is sufficiently great. In the absence of cholesterol, it is
not. The second and third terms in the above express the tendency of cholesterol
to increase the order of the saturated chains, provided that order is
not too large; i.e. cholesterol increases the order in the liquid phase but
decreases it in the gel phase \cite{vist90}. The attractive term,
proportional to $cs\delta$, is
crucial as it causes the addition of cholesterol to increase the chain order 
which thereby increases the repulsion between saturated and
unsaturated lipids so that they separate. Collecting the separate terms
of eq. (\ref{fliq1}) we have
\begin{eqnarray}
\label{fliq2}
{\tilde f}_{liq}(T,c,s,\delta)&=&  c\ln c+s\ln s+u\ln u\nonumber\\ 
&+&
J_{ss}s^2[k_1(\delta-1)^2+(\delta-1)^4]\nonumber \\
                     &+&
 J_{us}us\delta-J_{cs}cs(\delta-k_2\delta^2).
\end{eqnarray}
 Because the  order parameter is not controlled externally,
its value as a function of composition, $\delta_{liq}(T,c,s)$, is determined from the condition
that it minimize the free energy. Then the Helmholtz free energy per
particle of the liquid phase, $f_{liq}(T,c,s)$ is obtained from ${\tilde
f}_{liq}(T,c,s,\delta)$ according to 
\begin{equation}
f_{liq}(T,c,s)\equiv{\tilde f}_{liq}(T,c,s,\delta_{liq}(T,c,s)).
\end{equation} 
Two-phase coexistence is found in the usual way. Of the four unknowns,
the two independent compositions in the two coexisting phases, three are
determined by the conditions of equality of two independent chemical
potentials and of the Gibbs free energy per particle, $g$, which is essentially the
surface tension:
\begin{eqnarray}
\label{muc}
    \mu_c(T,c_1,s_1)&=&\mu_c(T,c_2,s_2),\\
    \mu_s(T,c_1,s_1)&=&\mu_s(T,c_2,s_2),\\
    g(T,c_1,s_1)&=&g(T,c_2,s_2),\qquad{\rm where}\\
    \mu_c&=&\frac{\partial f_{liq}}{\partial c}, \\ 
    \mu_s&=&\frac{\partial f_{liq}}{\partial s},\\
\label{gibbs}
     g&=&f_{liq}-\mu_cc-\mu_ss.
\end{eqnarray}
Because one composition is undetermined, there can be a region of concentrations
over which  two-phase
coexistence occurs.

\begin{figure}
   \begin{center}
      \includegraphics*[width=5.25in]{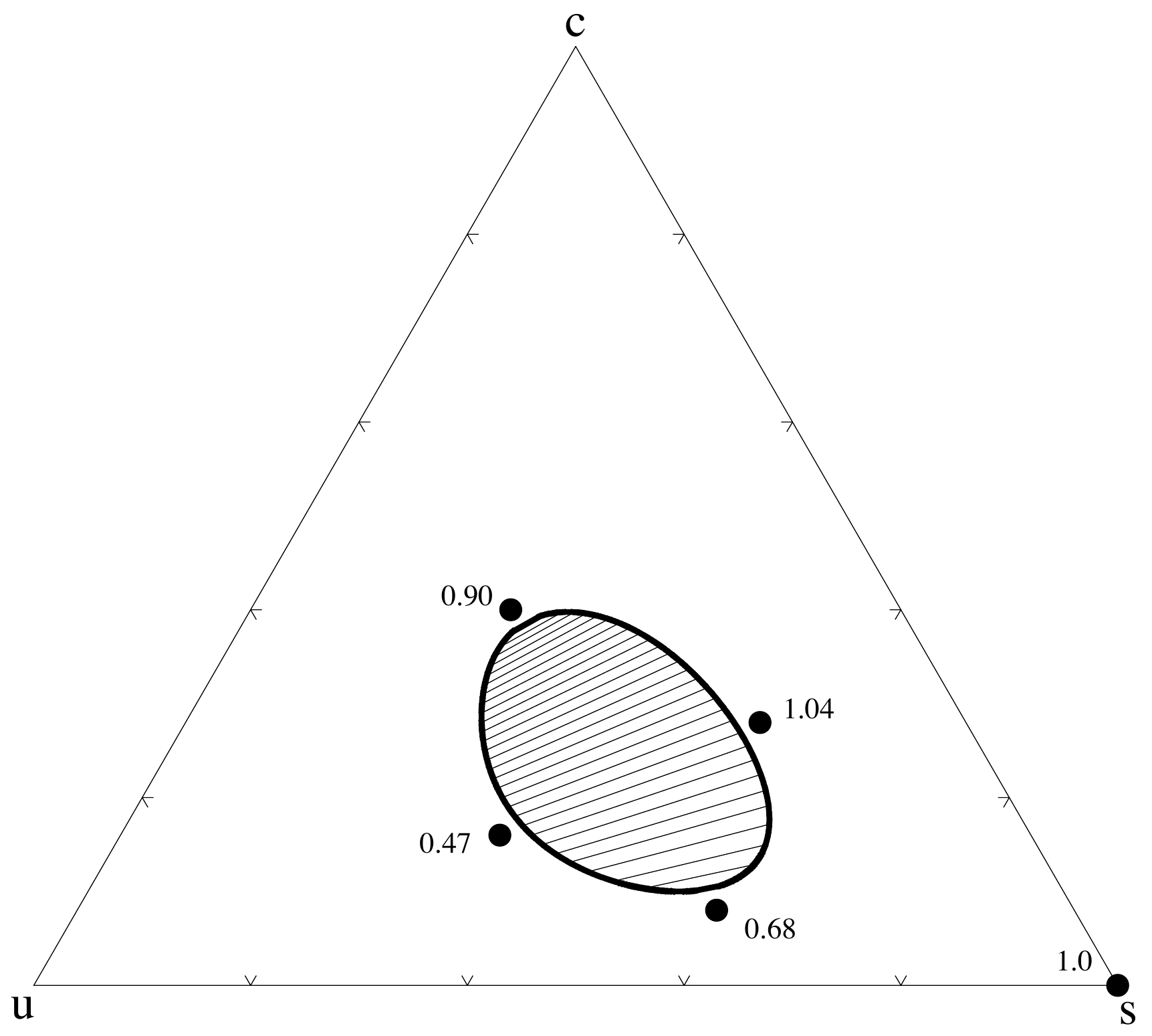}
      \caption{}
      \label{fig1}
   \end{center}
\end{figure}

A phase diagram which results from this theory is shown in Fig. 
\ref{fig1}. It reproduces the kind of phase diagram 
found in the diPhyPC, DPPC, cholesterol system. We have indicated values 
of the order parameter at several concentrations. As expected, it is large 
in the regions in which the concentration of the saturated
lipid is
large, and tends to increase with the concentration of cholesterol due to its ordering 
effect.

We now consider the system to be at a temperature at which it could 
exhibit a gel phase in addition to liquid phases. For the free energy of 
the gel phase we write
\begin{eqnarray}
\label{gel}
{\tilde f}_{gel}(T,c,s,\delta)&=&{\tilde f}_{mix}+{\tilde
f}_{chain,gel}+{\tilde f}_{int,gel},\nonumber \\
        {\tilde f}_{mix}&=& c\ln c+s\ln s+u\ln u,\nonumber \\
        {\tilde f}_{chain,gel}&=&J_{ss}s^2[k_1(\delta-2)^2+(\delta-2)^4 +k_3], \nonumber \\
        {\tilde f}_{int,gel}&=&J^{\prime}_{us}us\delta+J^{\prime}_{cs}cs\delta.
        \end{eqnarray}
        We note the following. The single minimum of the free energy in 
the gel phase has been set to occur at $\delta=2$ when the system consists 
only of the saturated lipid. In such a system the free energy of the gel 
phase exceeds that of the liquid phase by $k_3J_{ss}$, hence $k_3$ is 
proportional to $T-T^*$, with $T^*$ the liquid-gel transition temperature 
in the pure system. By changing $k_3$ from positive to negative values,
we induce a liquid to gel transition in our system. 
The interaction strengths $J^{\prime}_{us}$ and 
$J^{\prime}_{cs}$ are positive so that the addition of cholesterol, as 
well as unsaturated lipid, reduces the order of the saturated lipid in 
the gel phase \cite{vist90}. The value of the order parameter, 
$\delta_{gel}(T,c,s)$, is determined by minimization of the above free 
energy, and the Helmholtz free energy in this phase is
\begin{equation}
f_{gel}(T,c,s)={\tilde f}_{gel}(T,c,s,\delta_{gel}(T,c,s)).
\end{equation}
        
\begin{figure}
   \begin{center}
      \includegraphics*[width=5.25in]{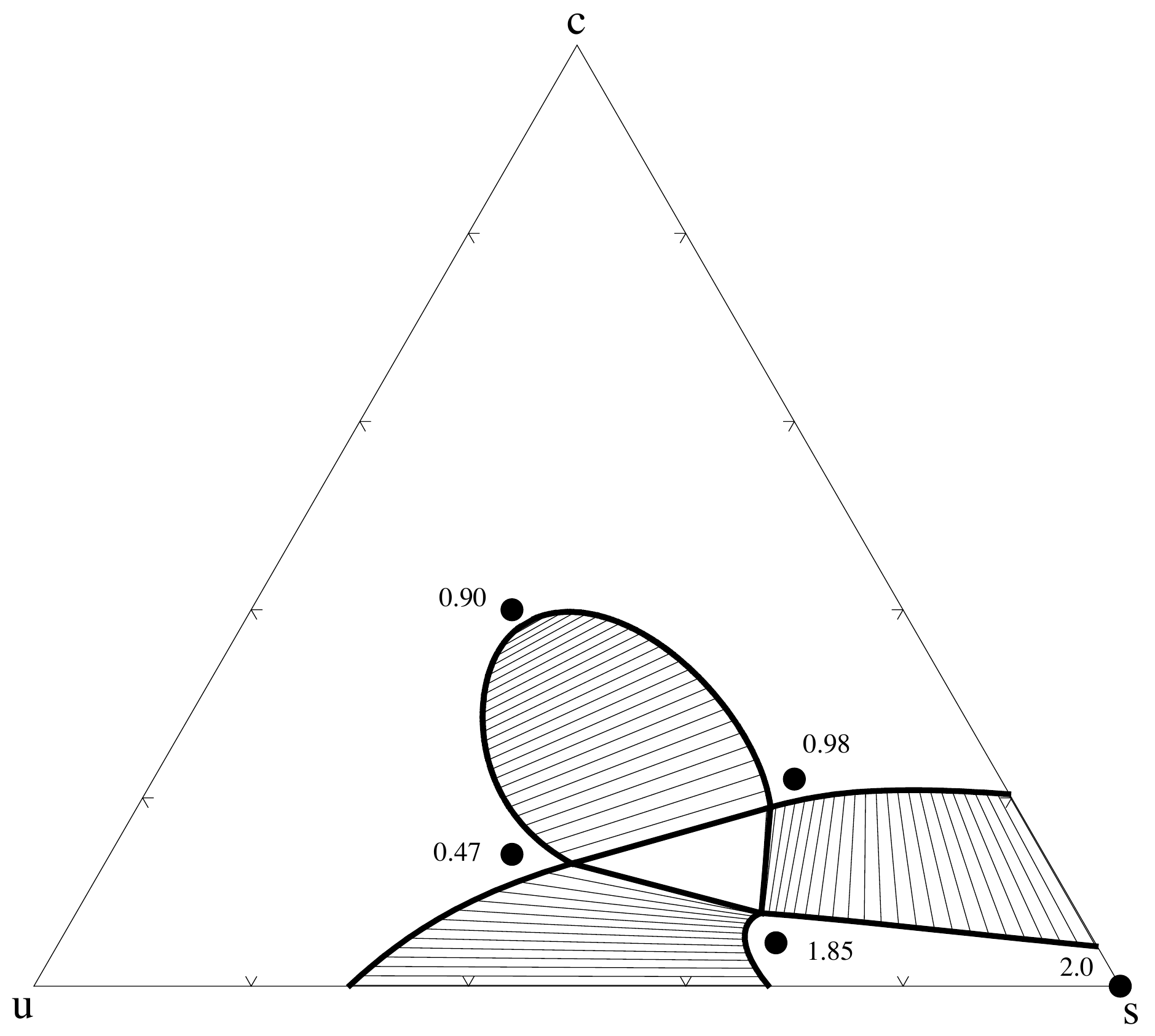}
      \caption{}
      \label{fig2}
   \end{center}
\end{figure}

The free energy of the entire system is now
        \begin{equation}
        f(T,c,s)=\min[f_{liq}(T,c,s),f_{gel}(T,c,s)],
        \end{equation}
and phase coexistence is again found by the conditions of the 
equality of two chemical potentials and of the Gibbs potential as in
eqs. (\ref{muc}) to (\ref{gibbs}) but with $f_{liq}(T,c,s)$ replaced by
$f(T,c,s)$ above. A phase 
diagram which results when the temperature is below that of the gel
transition of the saturated lipid is shown in Fig. \ref{fig2}. Values of
the order parameter at various concentrations are shown. The phase diagram shows
all the features of most of those observed experimentally. In
particular, there are two liquid phases and a gel. Tie lines between the
liquid phases indicate a greater presence of cholesterol in one liquid
than the other. The values of the order parameter in the coexisting
phases justify the designation of ``liquid ordered'' for the one and
``liquid disordered'' for the other. The order is greatest in the gel
phase of course.  Little cholesterol is
needed to destabilize this phase.

\section{Discussion}

We have presented a phenomenological theory of ternary mixtures of
cholesterol, a high-melting temperature (or saturated) lipid, and a low-melting
temperature (or unsaturated) lipid. It is capable of producing the usual phase diagrams
observed in this system in which there is a gel and two liquid phases,
and also the unusual phase diagram of the DiPhyPC, DPPC, cholesterol
system in which, for a range of temperatures,  
the ternary system exhibits phase separation even though none of the
binary systems do. The model highlights the interesting physics of these
systems; mixtures of high- and low-melting temperature lipids can phase
separate when the chains of the former are sufficiently well ordered, as
occurs in the gel phase. This is simply a result of packing
constraints. In the liquid phase, the chains of the
high-melting temperature lipid are not sufficiently ordered to induce
phase separation from the more disordered low-melting temperature
lipids. However the addition of cholesterol tends to order the chains of
the former until, with sufficient cholesterol, the chains have
sufficient order to bring about phase separation. 
This repulsion arising from packing constraints gives an explicit
origin to the unknown repulsion between low-melting temperature lipids
and the complexes of 
Radhakrishnan and McConnell \cite{radhakrishnan05}. However we do not 
introduce complexes
with a definite stoichiometry, but rather consider only the compositions
of the components of the  ternary system. We also note that, because our  theory is able to
describe the gel as well as the liquid phases, it is able to make 
clear that the existence of two liquid phases has little to do with
that of the gel. The gel phase simply occupies regions of the phase
diagram which might otherwise be occupied by liquid phases.  

The theory is sufficiently flexible to describe various possible
evolutions of the phase diagram with temperature, or equivalently, with 
the interaction strengths as they are generally inversely proportional
to temperature. With the parameters we have chosen in Figs. 1 and 2,
liquid-liquid phase separation remains even above the main chain
temperature of the saturated lipid. This can be traced to the large
repulsion, $J_{us}$, between saturated and unsaturated lipids which
might be expected given the bulky tails of diPhyPC. Were the magnitude
of this repulsion to be reduced, liquid-liquid coexistence would only be
observed below the main chain transition of the saturated lipid. 
As a second example, we note that the model is easily augmented to
permit the liquid-liquid coexistence to extend to the binary
cholesterol, unsaturated lipid system as is observed in some systems
\cite{almeida05}. 

Because of the attractive interaction between cholesterol and saturated
lipids in the liquid phase, the concentration of cholesterol is greater
in one of the liquid phases than the other. This agrees with some
experiments \cite{veatch04,veatch06,veatch07}, but not others
\cite{veatch04,ayuyan08}. Due to the preference of cholesterol for the
saturated over the unsaturated lipids, it enhances their 
phase separation; {\em i.e it increases} the temperature at which such
separation occurs \cite{prigogine54}. 
This is clear from the phase diagram of Fig. 1. The binary
lipid system is above its temperature of phase separation, but as cholesterol
is added, that temperature increases until it exceeds the temperature
for which the diagram is drawn. It follows from this that cholesterol
is not expected to migrate to the interface between lo and ld phases and
therefore is not expected to reduce the line tension between these
phases. Thus, if one wants to think of the existence of ``rafts'' as
indicating the formation of some sort of microemulsion, then one cannot
attribute to cholesterol the role of a line active agent which brings it
about. Rather one must assign such a role to a  
protein \cite{hancock06}, like N-Ras, which has two
anchors, one of which prefers the lo environment, and the other the
ld environment \cite{nicolini05}. 

Finally we note that, while we have referred to  the non-lipid component in the ternary system as cholesterol, it could equally be another molecule, such as a  protein. The crucial ingredient of our theory is that the non-lipid component interact with the saturated lipid in such a way as to increase its order parameter. If that be the case, then such a ternary system should exhibit a similar phase diagram to the one we have calculated here. Analogous conclusions also apply to a system consisting of such a protein {\em and} cholesterol and the unsaturated and saturated lipids.

\section{Acknowledgments} 
We thank John Cahn, Marcus Collins, and
Sarah Keller for useful conversations, and the National Science Foundation
for support under
grant no. DMR-0503752.
\clearpage

\clearpage
\section{Figure Captions}
\begin{itemize}
\item[]{Figure 1}
Phase diagram of the ternary system at a temperature above the main
chain transition of the saturated lipid. Values of the order parameter are shown at four
composition marked by dots. The values of the parameters are as follows:
 $J_{ss} = 1.0, k_1 = 1.0, J_{us} = 1.8, J_{cs} = 2.4, k_2 = 0.21.$
\item[]{Figure 2}
Phase diagram of the ternary system as a temperature below the main
chain transition of the saturated lipid. Values of the order parameter are shown at four
compositions shown by dots. The values of the parameters are as follows:
$J_{ss} = 1.0, k_1 = 1.0, J_{us} = 1.8, J_{cs} = 2.4, k_2 = 0.21, k_3 =
-0.25, J^{\prime}_{us}
= 0.7, J^{\prime}_{cs} = 0.0.$
\end{itemize}
\clearpage


%

\clearpage
\clearpage

\clearpage
\end{document}